# Towards Enhanced Analysis of Lung Cancer Lesions in EBUS-TBNA: A Semi-Supervised Video Object Detection Method

Jyun-An Lin[1]*, Ching-Kai Lin[1,2,3,4], Yun-Chien Cheng[1]*,

[1] Department of Mechanical Engineering, College of Engineering, National Yang Ming Chiao Tung University, Hsin-Chu, Taiwan

[2] Department of Medicine, National Taiwan University Cancer Center, Taipei, Taiwan

[3] Department of Internal Medicine, National Taiwan University Hospital, Taipei, Taiwan

[4] Department of Internal Medicine, National Taiwan University Hsin-Chu Hospital, Hsin-Chu, Taiwan

*Corresponding author: andy.en11@nycu.edu.tw, yccheng@nycu.edu.tw

## Abstract

This study aims to establish a computer-aided diagnostic system for lung lesions using endobronchial ultrasound (EBUS) to assist physicians in identifying lesion areas. During EBUS-transbronchial needle aspiration (EBUS-TBNA) procedures, physicians rely on grayscale ultrasound images to determine the location of lesions. However, these images often contain significant noise and can be influenced by surrounding tissues or blood vessels, making identification challenging. Previous research has lacked the application of object detection models to EBUS-TBNA, and there has been no well-defined solution for the lack of annotated data in the EBUS-TBNA dataset. In related studies on ultrasound images, although models have been successful in capturing target regions for their respective tasks, their training and predictions have been based on two-dimensional images, limiting their ability to leverage temporal features for improved predictions. This study introduces a three-dimensional video-based object detection model. It first generates a set of improved queries using a diffusion model, then captures temporal correlations through an attention mechanism. A filtering mechanism selects relevant information from previous frames to pass to the current frame. Subsequently, a teacher-student model training approach is employed to further optimize the model using unlabeled data. By incorporating various data augmentation and feature alignment, the model gains robustness against interference. Test results demonstrate that this model, which captures spatiotemporal information and employs semi-supervised learning methods, achieves an Average Precision (AP) of 48.7 on the test dataset, outperforming other models. It also achieves an Average Recall (AR) of 79.2, significantly leading over existing models.

Keywords: Deep learning, Lung cancer, Vision Transformer, Endobronchial ultrasound-guided transbronchial needle aspiration, Mediastinal lesions, Video object detection

## 1. Introduction

Lung cancer is the leading cause of cancer incidence and mortality [1], and early diagnosis with accurate staging is crucial for subsequent treatment planning [2]. In the diagnostic process, live biopsy surgeries are commonly employed. Endobronchial ultrasound-guided transbronchial needle aspiration (EBUS-TBNA) is a relatively new minimally invasive technique widely used for sampling lesions in the mediastinal cavity and lung hilum. Previous studies have confirmed its high accuracy in diagnosing and staging malignant lung tumors [3-6]. Within the limited anesthesia time, being able to accurately locate and sample lesions on ultrasound videos can effectively improve the efficiency and accuracy of physicians during surgery, thereby reducing the risks associated with the procedure.

EBUS-TBNA imaging includes three modalities: grayscale, Doppler, and elastography. Grayscale is used to locate lesions and perform initial analysis of lesion features [7]. Doppler detects blood flow signals around lymph nodes, confirming the condition of lymph nodes and other tissues [8]. Elastography measures the rigidity of mediastinal and hilar lymph nodes to further confirm whether they are malignant lesions [9]. Grayscale imaging is a prerequisite for subsequent imaging analysis. Although grayscale imaging can reveal the location of lesions, it is sometimes affected by interference such as blood vessels and noise. Bronchoscopic physicians also require more learning experience to familiarize themselves with these imaging features. Therefore, efficiently analyzing the features of grayscale imaging to quickly identify suitable lesions for subsequent benign-malignant interpretation is essential.

Previous studies have applied deep learning to EBUS-TBNA [10-12], but they primarily focused on classification, which does not align with our objectives. To assist doctors in locating lesions, past research has used deep learning-based object detection models on ultrasound images [13-17] to generate prediction boxes. Among these models, since transformers are the current state-of-the-art architecture, newer models for both EBUS-TBNA and other ultrasound images are based on modifications of the transformer architecture [10][17]. The DETR design by Tang's team is also based on the transformer architecture [17]. They used this model to predict spinal curvature and vertebral levels, and compared to other research architectures, its unique attention mechanism better captures global information to improve performance. However, the above-mentioned methods for ultrasound images only capture spatial information from the current frame and cannot obtain temporal information. During surgical procedures, the position of lesions changes over time. We believe that these two-dimensional image-based model architectures, if unable to effectively capture temporal information, will lead to errors in identifying dynamic EBUS-TBNA images. Additionally, the annotated data in our dataset requires doctors to review the entire video to determine the location, making it difficult to obtain annotations. The methods also do not address the issue of handling limited annotated data.

To address the challenge of not capturing temporal information on dynamic EBUS-TBNA videos, we aim to design a video object detection model based on three-dimensional video data. Most newer methods are based on modifying DETR [18-21] to leverage its unique attention mechanism to capture temporal correlations and thus obtain temporal information. For example, Si's team designed the DiffusionVID architecture [21], which uses information from other frames to perform attention calculations, then uses Euclidean distance to filter queries, using these queries for subsequent attention calculations to obtain the result. This approach utilizes temporal information. A distinctive aspect is that they incorporate the concept of diffusion models, removing redundant boxes through denoising and gradually refining them to the optimal boxes. We aim to design an video object detection model for the EBUS-TBNA dataset based on these methods. During EBUS-TBNA surgery, even slight movements of the probe can cause significant changes in the position of lesions in the ultrasound images. Therefore, we plan to use an attention mechanism to capture temporal information and incorporate a filtering mechanism to filter out areas with large positional changes, preventing unsuitable temporal information from causing prediction interference.

Moreover, to address the issue of the labor cost associated with annotation data, we aim to efficiently utilize unlabeled data. Previous research has applied semi-supervised methods to object detection models [22-24]. For instance, the Semi-DETR designed by Zhang's team [22] uses different data augmentations for the pre-trained teacher and student models and aligns the queries to achieve cross-view query consistency. It combines both many-to-many and one-to-one matching methods. Finally, it incorporates the Mean-Teacher semi-supervised learning method to utilize unlabeled data, thereby improving the accuracy of the student model. Existing semi-supervised learning studies mostly use a teacher-student model pair to improve the quality of pseudo-labels, increase the model's robustness, and thereby more effectively utilize unlabeled data. Many of these studies employ methods for contrastive learning with added noise and aligning the outputs or features of the student and teacher models. In EBUS-TBNA data, the predictions of the teacher model are more likely to be influenced by factors like blood vessels, tissues, etc., leading to potential misjudgments of lesion areas. Features prone to misjudgment are expected to have lower confidence scores. Therefore, we aim to align the student and teacher models to calculate the loss and weight the results based on confidence scores, increasing the proportion of correct information transmitted. Additionally, we intend to introduce contrastive learning with noise commonly found in EBUS-TBNA data, making the student model less susceptible to noise effects.

In the mentioned literature, it is evident that capturing temporal features in video data leads to stronger performance. Applying semi-supervised methods to object detection models effectively leverages unlabeled data for improved model performance. However, there is currently no model well-suited for EBUS-TBNA data in the literature. Therefore, we aim to reference the studies to design a semi-supervised object detection model based on video data.

While the application of object detection models in ultrasound images has validated the performance of deep learning architectures in analyzing EBUS-TBNA videos, we have identified the following challenges in existing techniques as follows:
(1) No research has proposed effective methods to simultaneously acquire temporal information from ultrasound videos and integrate it with EBUS-TBNA videos.
(2) Some model architectures proposed in previous studies do not address the issue of limited annotated data in the dataset. Utilizing these models may result in degraded performance due to insufficient annotated data.

To enable the model to capture temporal information from dynamic EBUS-TBNA videos and apply it to clinical settings, this study has improved and designed a three-dimensional model based on the Detection Transformer (DETR). In recent years, with the rise of transformer architectures, DETR has become a widely used deep learning framework for computer vision object detection tasks. Its powerful self-attention mechanism has demonstrated better accuracy than CNNs in various public datasets [25][26]. However, subsequent video object detection models derived from DETR, while exhibiting good performance, require large amounts of data for training due to their numerous parameters. This makes it challenging to apply DETR-based modified video object detection models to datasets with limited annotated data. To address this issue, we applied a semi-supervised learning approach to effectively utilize unlabeled data. Additionally, by observing EBUS-TBNA videos, we further tailored our approach to the characteristics of this data. The entire training framework is referred to as DEBUS.

The contributions of this study are summarized as follows:
(1) This study aims to design a computer-assisted system capable of outlining lesions for reference by medical professionals.
(2) The proposed SVDETR model leverages the temporal sequence information in videos to achieve higher performance.
(3) This study introduces random masking in image data, requiring the model to use temporal features to determine the current target location.
(4) By using semi-supervised learning to align different features between the teacher and student models, this study effectively utilizes unlabeled data and reduces the model's dependence on annotated datasets. The results of aligning different features are compared.
(5) This study incorporates a diffusion model module, enabling the model to generate a better set of queries

based on different images, which helps the model identify lesion locations and improves training.

In conclusion, the goal of this research is to develop a computer-assisted diagnostic system. Through the proposed DEBUS framework, higher performance is achieved compared to existing methods, while also saving manpower and time costs in clinical applications. This system aims to provide real-time confidence scores and prediction boxes to assist physicians in analyzing the location of lesions in the thoracic cavity, reducing the sampling time during anesthesia, and ultimately improving the diagnosis of lung cancer staging.

## 2. Related Work

### 2.1. DEtection TRansformer (DETR)

DETR is an object detection model based on the Transformer architecture [27]. Its core idea is to utilize the attention mechanism (Multi-head Attention) to capture information from the entire image in a global context. The model initially extracts features from the image using a convolutional neural network and introduces positional encoding to preserve location information. The reduced feature map, combined with spatial position encoding, is then input into the encoder. The encoder employs a self-attention mechanism to weight different positions in the input, capturing global information. Results of the encoder is the features encoded for N objects.

Next is the decoder, where each decoder takes two inputs: an object query and the output of the encoder, used to decode information for N objects. Each layer of the decoder calculates losses and includes learnable positional encoding. Finally, two feed-forward networks generate predicted detection boxes and categories.

Compared to other deep learning architectures, the attention mechanism can capture long-term relationships at different positions and assign different weights based on the importance of different parts of the data. Additionally, when processing images, it can globally capture information. Therefore, it is widely popular in image classification or object detection tasks.

### 2.2. Semi-supervised Learning Method

Semi-supervised learning is a machine learning approach that improves model performance by effectively utilizing both labeled and unlabeled datasets. Literature has shown that the Mean-teacher method[28] in semi-supervised learning is particularly suitable for large models like transformers[29]. In practice, labeled data is labor-intensive and time-consuming to obtain, whereas unlabeled data is relatively easy to acquire. Semi-supervised learning can efficiently use unlabeled data to enhance the model's generalization performance.

While label propagation is a common semi-supervised learning method, some models use feature alignment to enable the student model to learn from the teacher model. Initially, the model undergoes supervised learning using labeled data to learn how to predict labels. Then, it aligns the outputs of various modules on differently augmented unlabeled data, using both labeled and unlabeled data for further training. This allows the model to learn more about data structures and representations from the unlabeled data. The Zhang team applied different data augmentation effects to unlabeled data and fed them into the student and teacher models.[22] By calculating the loss function based on the differences in their outputs, they effectively utilized unlabeled data. This approach demonstrated significant effects in semi-supervised learning on public datasets.

### 2.3. Diffusion Model

A diffusion model is a type of generative model that generates data by progressively adding noise and learning the reverse process. The fundamental principle involves two processes: the forward diffusion process and the reverse generation process [30].

(1) Forward Diffusion Process: Starting from the original data, random noise is gradually added to the data, making it increasingly blurred and unstructured. This process can be viewed as a Markov chain, where a small amount of Gaussian noise is added to the data at each step until the data becomes pure noise.

$$q(x_t|x_{t-1}) = N(x_t; \sqrt{1-\beta_t}x_{t-1}, \beta_t I) \quad (1)$$

(2) Reverse Generation Process: The diffusion model learns the reverse of the forward process, starting from pure noise and gradually removing the noise to recover the original data. The goal of training is to learn how to estimate and remove the noise from each step's noisy data, making it increasingly close to the original data distribution.

$$P(x_{t-1}|x_t) = N(x_{t-1}; \mu_0(x_t, t), \Sigma_0(x_t, t)) \quad (2)$$

Additionally, there have been other studies proposing improvements to diffusion models by adjusting the noise addition methods [31]. In other object detection tasks, related research has applied this method to enhance model performance and increase generalizability [21][32], achieving good results on public datasets. This demonstrates the strong potential for further development of diffusion models.

## 3. Material and Methods

### 3.1. EBUS-TBNA Dataset

This study collected data from 150 patients with mediastinal or hilar lesions who underwent EBUS-TBNA at the Chest Department of NTU Cancer Center from November 2019 to April 2021. The video recordings of the examinations were stored using a medical imaging recording system. The dataset includes 330 video segments of lesions, totaling 1183 annotated images and 5004 unannotated images, all of which are grayscale images. The average duration of annotated video segments is 3.8 seconds. For data partitioning, the study designated data collected after November 27, 2020, as the test dataset. Data collected before November 27, 2020, were randomly split by patient into training and validation datasets. The training set contains 798 annotated images and 5004 unannotated images, the validation set contains 171 images, and the test set contains 214 images. The study was approved by the National Taiwan University Cancer Center Institutional Review Board (IRB #202105105RIND).

### 3.2. Experimental Procedure

The experimental process is shown in Figure 1. This study will first collect data and divide it by patient into training, validation, and test sets. Then, data preprocessing and augmentation will be performed. Random masking data augmentation will be applied to the training set, requiring the model to learn how to locate the target position using temporal information. The performance differences between models with different ratios of random masking will be compared to verify the effectiveness of random masking data augmentation. For model design, the study will first refer to previous research on video object detection models and design a DETR-based video object detection model. Next, a semi-supervised learning method will be designed to further optimize the model. Finally, multiple tests will be conducted, including:

(1) The effect of applying different numbers of image frames on the model's results.
(2) The impact of different noise strategies in diffusion models on the model.
(3) Testing the effectiveness of the random masking enhanced model in extracting temporal information.
(4) The impact of different numbers of filtering mechanisms on the model.
(5) Testing the model before and after applying the semi-supervised learning method.
(6) Comparing the test results of various models.
(7) Testing the model running speed.

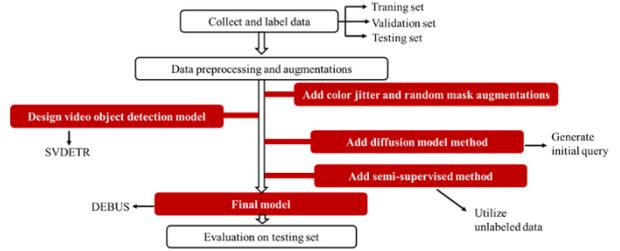

Figure 1. Experiment process

### 3.3. Data Preprocessing and Augmentation

To reduce the model's computational load, this study crops the EBUS-TBNA images from the original image (1920×1200) to a fixed size and position. Since the lesion position in EBUS-TBNA videos moves over time and the lesions in each frame are correlated, this study segments the video into short clips of five seconds each and extracts frames at one-second intervals, then stacks them into a 3D dynamic image (clip) as input. To effectively utilize the annotated video segments, our model architecture is designed to use five-second videos along with their annotated data, allowing the model to receive more video information. Padding is used to enable the model to capture spatial information, so even in the absence of temporal information, good results can still be achieved.

For data augmentation, common techniques such as random scaling, random flipping, random padding, etc., are used, along with additional augmentations as shown in Figure 2, such as the specially designed ColorJitter. This can confuse the color of vascular tissue and lesions, forcing the model to rely on shape and surrounding information to identify the lesion location. Additionally, since we use image clips, we add random masks to some frames of each clip, requiring the model to use temporal information to find the current lesion location, further enhancing the model's ability to extract temporal features.

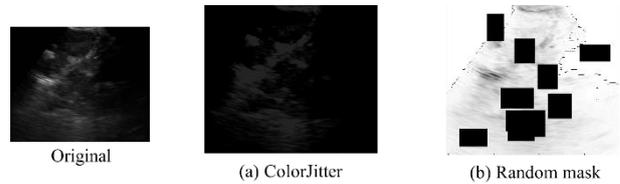

Figure 2. Data augmentation diagram

### 3.4. Model Overview

The model designed in this study is called DEBUS, as shown in Figure 3. DEBUS is built upon the architecture of DETR. It includes a SVDETR component capable of capturing temporal information in images. Additionally, through the teacher-student model in semi-supervised learning, the model is further optimized using unlabeled datasets. The following sections will introduce these two components separately.

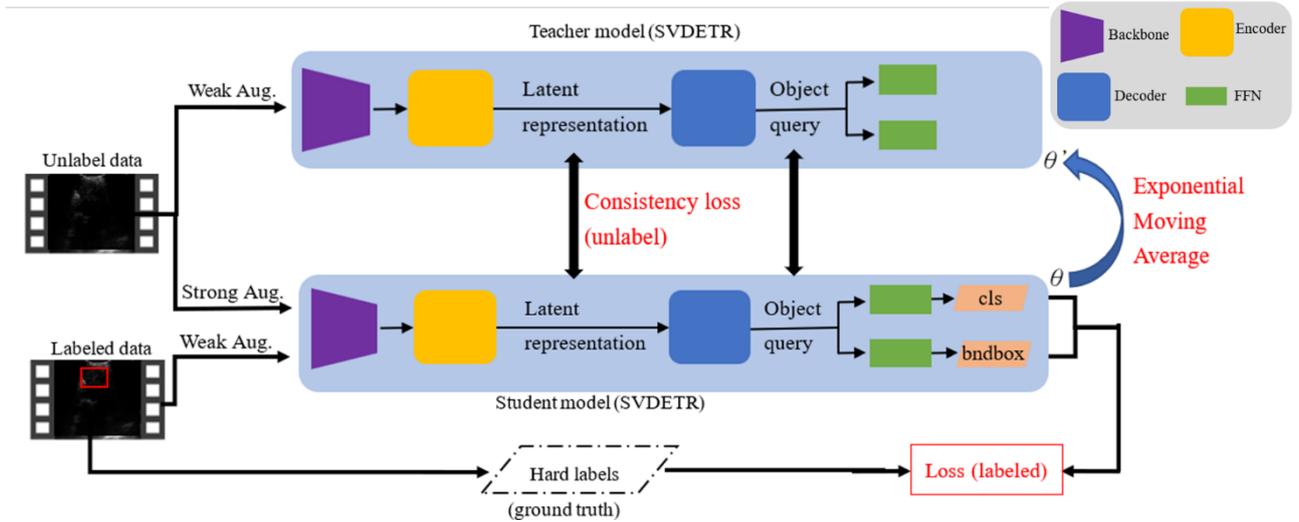

Figure 3. Overview of DEBUS

*3.4.1. SVDETR (Short time video detection transformer)*

The SVDETR model, as depicted in Figure 4, is designed in this study to capture temporal information in images. We employ a two-dimensional CNN to extract features from each frame of the image. These features are then separately inputted into the encoder of DETR for encoding. The encoded results of each frame are then fed into the decoder, where an attention mechanism calculates attention with the previous frame's image, effectively transmitting past information to the last frame and capturing temporal information.

In this architecture, we maintain the design of DETR, with six layers in the encoder and two layers in the decoder, both sharing weights. To ensure accurate detection of lesions in each frame, a loss function is calculated for the output of the decoder for every frame. Additionally, as the model progressively captures temporal information, the initial frames may lack accuracy. To address this, we use a diffusion generative model to generate a set of initial queries for the first frame prediction.

Due to the significant changes in ultrasound images caused by variations and movements of the ultrasound probe during EBUS-TBNA procedures, we retain the frames with the highest attention scores in the attention mechanism propagation process of relevant frames. This information is then transmitted together to assist the prediction of the current frame. To accommodate the grayscale image length of EBUS-TBNA, we design the model to take five frames as input at a time. To validate the effectiveness of the added modules in EBUS-TBNA, we will separately test the performance of different diffusion generative strategy, and attention filtering mechanisms.

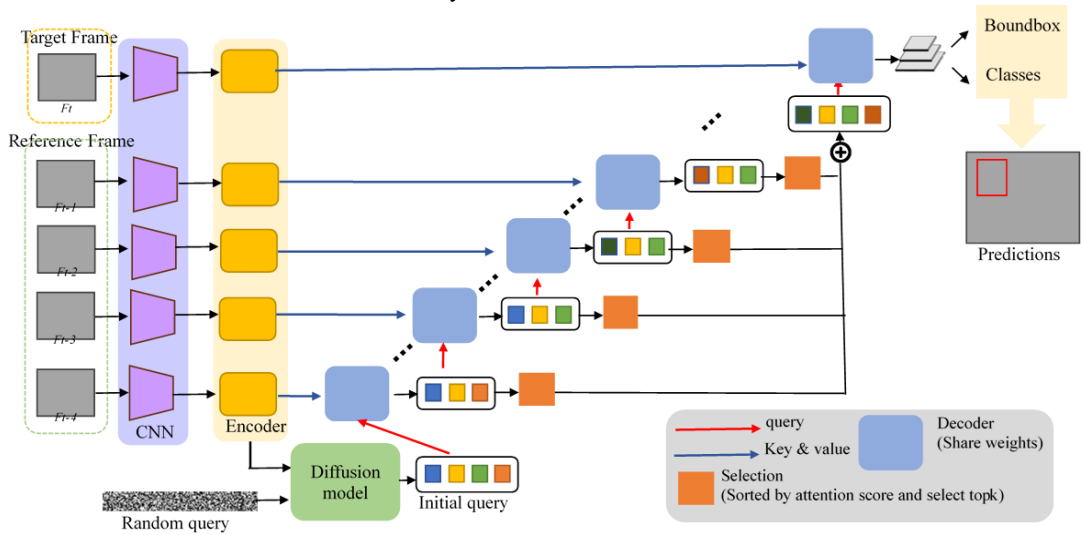

Figure 4. SVDETR model architecture diagram

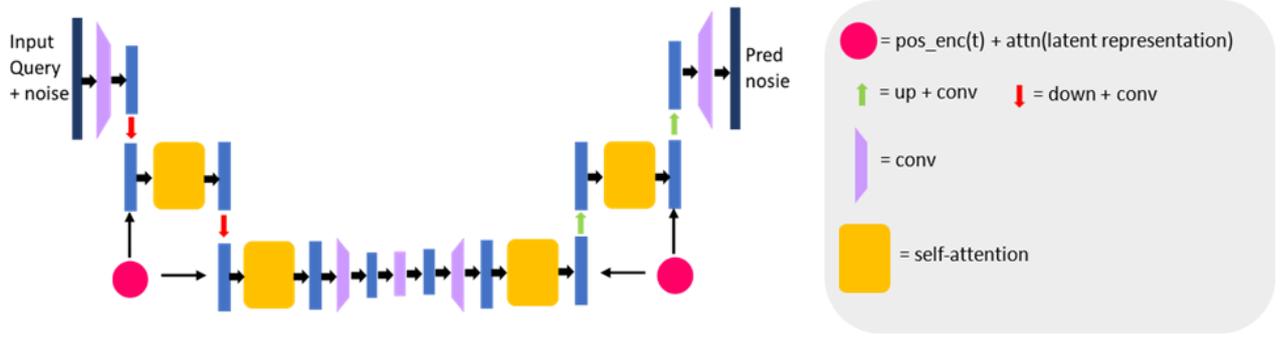

Figure 5. Diffusion model architecture diagram

### 3.4.2. Diffusion generative model

To implement the diffusion model's denoising and query generation from noisy images, we designed a Unet architecture (see Figure 5), with an illustration of training and validation shown in Figure 6. During training, the diffusion model gradually adds noise to train the model, while during validation, it gradually restores the noise back to the original image. Therefore, in this study, during the training phase, the input of the diffusion model consists of the time steps, encoder outputs, and the query produced from the last frame with added noise. The model estimates a set of noise based on the time steps and encoder outputs, calculates the difference from the actual added noise values, and updates parameters accordingly. During the validation and testing phases, the model gradually restores the noise based on the time steps and encoder outputs to generate a set of queries that match the first frame, serving as initial queries for subsequent use.

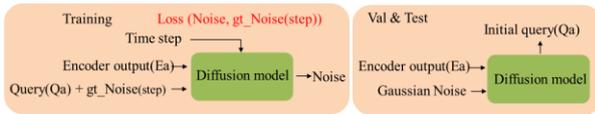

Figure 6. Diagram of train and test diffusion model

### 3.4.2. Semi-supervised method(Teacher student model)

Through our experiments, we have verified that video-based object detection models are highly beneficial for EBUS-TBNA data. However, they still struggle to effectively utilize other unlabeled data. Moreover, since such models require continuous annotated data as input, it further increases the difficulty of annotation.

To enhance the accuracy and robustness of the model, we introduced the mean-teacher framework into the SVDETR model. This framework mainly consists of a teacher model and a student model, both incorporating different data augmentation techniques. While the architectures of both models are identical, their parameter update methods differ. Below are the details of the approach:

(1) Architecture Overview: As shown in Figure 3, both the teacher model and the student model consist of the same backbone, encoder, and decoder. The student model handles both annotated and unannotated data, while the teacher model is used solely for inference on unannotated data.

(2) Parameter Alignment: To maintain consistency between the teacher model and the student model, we use Exponential Moving Average (EMA) to align the parameters of the teacher model. In each training step, the parameters of the teacher model are updated according to the following formula:

$$\theta_{teacher} = \alpha\theta_{teacher} + (1-\alpha)\theta_{student} \quad (3)$$

(3) Consistency Loss: We introduce a consistency loss between the student model and the teacher model, calculated on unlabeled data, to ensure that their predictions are similar.

(4) Loss Function Integration: The final loss function of the student model consists of the classification loss and bounding box regression loss of annotated data, along with the consistency loss of unlabeled data.

$$L = L_{labeled} + L_{consistency} \quad (4)$$

Through the above methods, we effectively aligned the modules of the student and teacher models and utilized unlabeled data for semi-supervised learning, significantly improving the model's ability to detect lesions in EBUS-TBNA images.

### 3.5. Equipment

This study utilized a server with ASUS Z790-A GAMING WIFI 6E, powered by Intel I9-13900K, and equipped with MSI RTX 4090 GAMING X TRIO 24G.

## 4. Result

### 4.1. Evaluation Methods

For all object detection model architectures, this study will use AP and AR calculations as indicators to evaluate the model's performance. The AP calculation method involves first sorting the prediction results by confidence score and then checking the prediction results based on different confidence thresholds. If the prediction matches the ground truth, TP (True Positive) is incremented by 1; if there is no match, FP (False Positive) is incremented by 1. FN (False Negative) is the number of ground truth labels that were not detected. Precision is calculated as TP/(TP+FP) and Recall as TP/(TP+FN). After plotting the Precision-Recall curve, the area under the curve is the AP. The mAP is the average AP across different categories. Since our task only detects lesions, AP is equivalent to mAP. Additionally, AR is the average result based on different IoU thresholds.

As there are currently no publicly available EBUS-TBNA datasets, this study only uses data provided by NTU Cancer Center for testing.

### 4.2. Implement Detail

The hyperparameters used in training DEBUS in this study include an initial learning rate of 5e-7, the optimizer is Adam with Weight Decay Regularization (AdamW), the classification loss is Cross-entropy Loss, and the bounding box loss is a combination of L1 Loss and GIoU Loss. The parameters are updated every 4 data points. The total training epoch for DEBUS is 300, with a learning rate decay to 0.1 every 120 epochs.

### 4.3. Comparison of Different Models

This section aims to compare SVDETR with other object detection models. Table 1 shows the results of different models on the test set. From the results, it can be observed that SVDETR and DEBUS models proposed in this study have significant advantages in terms of prediction box accuracy. Under the evaluation standards of AP (Average Precision) and AR (Average Recall), both models outperform other image-based object detection models [27][33][34]. This indicates that the SVDETR and DEBUS architectures can effectively capture temporal features and utilize them to enhance prediction quality. Moreover, compared to other video-based object detection models [19][21][35-40], our models also demonstrate strong competitiveness in AP evaluation standards. Particularly, in terms of AR, the DEBUS model significantly surpasses all comparison models.

From the data in the table, it's evident that SVDETR and DEBUS models perform remarkably well on all metrics. Especially the DEBUS model, which not only achieves the highest values in AP and $AP_{75}$ but also significantly outperforms other models in terms of AR. These results confirm the effectiveness of our approach in capturing temporal information and enhancing prediction performance.

Table 1. Comparison of different models

| Model | AP | $AP_{50}$ | $AP_{75}$ | AR |
|---|---|---|---|---|
| YOLOX [33] | 41 | 82.7 | 35.6 | 49.2 |
| Faster-RCNN [34] | 40 | 83 | 31.7 | 50 |
| DETR [27] | 41.2 | 84.8 | 33 | 72.7 |
| DFF [35] | 20.5 | 58.1 | 10.7 | 35.8 |
| FGFA [36] | 32.3 | 76.3 | 18.3 | 42.9 |
| RDN [37] | 35.6 | 84.6 | 23.9 | 45.4 |
| MEGA [38] | 41.4 | 88.1 | 31.7 | 50.3 |
| DAFA [39] | 35.6 | 87.3 | 17.5 | 44.5 |
| TransVOD [19] | 41.1 | 81 | 35.4 | 59.1 |
| YOLOV [40] | 45.8 | **91.7** | 33.4 | 55.5 |
| DiffusionID [21] | 46.8 | 91.4 | 38.2 | 57.8 |
| SVDETR (Ours) | 47.5 | 89.4 | 42.2 | 78.1 |
| DEBUS (Ours) | **48.7** | 91.3 | **43.6** | **79.2** |

### 4.4. Influence of Random Masks Quantity on Model Performance

This section aims to explore the influence of the number of frames and the quantity of random masking on the performance of the model (SVDETR). Table 2. shows the results of the model on the test set under different settings. As can be seen from the table, the use of 10 random masks yields the best results. This indicates that the data augmentation method of random masking indeed enhances the model's ability to learn temporal features. However, when the number of masks is too high, the model's performance may decline due to a lack of spatial information.

Table 2. Comparison of SVDETR model results with different numbers of random masks

| Mask Num | AP | $AP_{50}$ | $AP_{75}$ | AR |
|---|---|---|---|---|
| - | 42.6 | 87.5 | 34 | 77.3 |
| 5 | 45.3 | **90.2** | 37.5 | **79.1** |
| 10 | **46.3** | 86.3 | **43.2** | 77.4 |
| 15 | 43.7 | 87.9 | 35.2 | 76.7 |

### 4.5. Comparison of Results with Different Diffusion Strategies Applied to the Model

This section aims to discuss the impact of different diffusion model strategies on the SVDETR model. Table 3 shows various diffusion strategies, including methods that introduce noise over time steps: (1) Linear and (2) Cosine. Despite research [31] confirming that using the Cosine method to add noise can make the noise more varied, this approach is more susceptible to noise interference. As shown in the table, the performance of the Cosine method is inferior to that of the Linear method. We speculate this is because the generated queries themselves already contain a certain amount of noise. When using the Cosine method, it may be difficult for the model to determine the origin of the noise, leading to decreased performance. In some cases, the effect of adding noise using the Cosine method is even worse than not adding noise at all.

From the results of adding noise with the Linear method, the model's average precision (AP) significantly improves. As seen in Figure 7, the initial queries generated by the diffusion model indeed help locate lesions, thereby confirming the effectiveness of the diffusion model in enhancing model performance.

Table 3. Comparison of different diffusion model strategies applied to the model

| Diffusion method | AP | $AP_{50}$ | $AP_{75}$ | AR |
|---|---|---|---|---|
| - | 46.3 | 86.3 | **43.2** | 77.4 |
| Linear [30] | **47.5** | **89.4** | 42.2 | **78.1** |
| Cosine [31] | 45.2 | 88.8 | 36.1 | 77.6 |

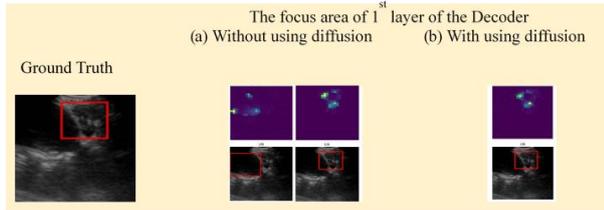

Figure 7. Visualizing diffusion model results

**4.6. Comparison of Results of Different Quantitative Screening Mechanisms**

This section aims to delve into the impact of different filtering mechanisms on the performance of image object detection models. The filtering mechanism here refers to preserving the query results that are more likely to be target objects from previous frames and uniformly inputting them in the current frame. From the results in Table 4, it can be observed that the influence of different filtering mechanisms on the model is not significant, with minimal differences between them. We believe this result may be due to the relatively small number of target objects in the study, making it difficult for the advantages of the filtering mechanism to be demonstrated in handling multi-target detection. Additionally, the original temporal propagation mechanism, by passing information between consecutive frames, is already capable of capturing dynamic changes, hence the limited improvement effect of the filtering mechanism. Furthermore, the filtering mechanism may not have fully considered various variables in the specific scene design, resulting in its limited ability to significantly enhance model performance in practical applications.

Table 4. Different quantity screening mechanism model result table

| Top K | AP | $AP_{50}$ | $AP_{75}$ | AR |
|---|---|---|---|---|
| - | 47.5 | 89.4 | 42.2 | 78.1 |
| 1 | 47.2 | 89.7 | 42.3 | 78.3 |
| 3 | 47.4 | 89.6 | 42.6 | 78.1 |
| 10 | 47.5 | 90 | 42.4 | 78 |
| 20 | 47.6 | 90.1 | 42.4 | 77.7 |

**4.7. Comparison of Semi-supervised Learning Applications in Video Object Detection Models**

This section aims to discuss the performance of semi-supervised learning methods in video object detection models and the differences in aligning various modules. Table 5 demonstrates that models incorporating semi-supervised learning achieve higher average precision (AP) values. Specifically, on the test set, there are significant improvements in AP (+1.2), $AP_{50}$ (+1.2), $AP_{75}$ (+1.4), and an increase in average recall (AR) (+1.1). This validates the effectiveness of semi-supervised learning for the EBUS-TBNA dataset, which has limited annotated data. Through this method, the model can learn spatial features of EBUS-TBNA data that lack labels.

Furthermore, we compared the effects of aligning different module positions and found that aligning the encoder and decoder yields better results. We speculate this is because the backbone represents high-level semantic features of the input data. Without important information extracted by the encoder to transform into global contextual information, it contains too much noise. Therefore, aligning the encoder's focused global context output and the decoder's output used to predict queries may reduce error messages and thus improve model performance. We believe there is still much room for further research in this area.

Table 5. Comparison of different features of teacher and student models aligned to the model

| Align | | | AP | $AP_{50}$ | $AP_{75}$ | AR |
|---|---|---|---|---|---|---|
| backbone | encoder | decoder | | | | |
| | | | 47.5 | 89.4 | 42.2 | 78.1 |
| V | | | 46.7 | 89.5 | 40.9 | 77.6 |
| | V | | 48.5 | 90.2 | **43.8** | 78.5 |
| | | V | 47.6 | 90.2 | 43.5 | 78.9 |
| V | V | | 47.9 | 90 | 42.4 | 78.3 |
| V | | V | 47.4 | 90.2 | 43.1 | 78.2 |
| V | V | V | 48.1 | 90 | 43.6 | **79.3** |
| | V | V | **48.7** | **90.6** | 43.6 | 79.2 |

## 4.8. Model Speed Testing

Achieving clinical applicability requires assessing the operational speed of the model, which is a critical metric. In clinical settings, continuous acquisition of image data occurs. Based on our model's architectural design, we can store results from the previous frame to facilitate computations using a single-layer decoder, without needing to recalculate previous information. Table 6 presents the speed test results for DEBUS, showing that with this approach, we achieve a recognition speed of 0.039 seconds per frame after the first frame using the diffusion model. This translates to approximately 25.64 frames per second, whereas our video runs at 24 frames per second. Therefore, our model achieves real-time prediction capability when not employing the diffusion model.

Table 6. Model running speed table (†: without diffusion model)

| Model | Params. | GFLOPs | Sec/clip | FPS |
|---|---|---|---|---|
| DEBUS | 37.60M | 295.03 | 0.77 | 1.30 |
| DEBUS† | 32.53M | 57.73 | 0.039 | 25.64 |

## 4.9. Experiment summary

In section 4.3, the SVDETR proposed in this study achieved the highest AP of 47.5 and AR of 78.1 on the test set. Section 4.4 compares the performance using different frame rates and random masking data augmentation. Section 4.5 presents and analyzes the effects of different diffusion model strategies on the model. In section 4.6, the differences in model performance with different numbers of filtering mechanisms are compared and discussed. In section 4.7, the semi-supervised learning method framework DEBUS is introduced, resulting in an increase in AP to 48.7 and AR to 79.2. Finally, in section 4.8 discussed the model's operational speed, ensuring DEBUS can provide classification information in real-time clinical settings.

## 5. Conclusion

In the current study, the DEBUS model proposed achieves a prediction box accuracy of 48.7 AP and 79.2 AR on the test set of EBUS-TBNA images. DEBUS integrates SVDETR, leveraging its excellent temporal feature extraction capabilities and the initial query generation capability of the diffusion model, which results in superior performance compared to other image object detection model architectures. Furthermore, the effective use of semi-supervised learning with unlabeled data further enhances DEBUS's performance. We also verified that the filtering mechanism does not significantly impact the EBUS-TBNA dataset.

There are areas for improvement in our model. Future enhancements could focus on refining the diffusion model in SVDETR to address the considerable noise in queries, possibly transitioning the query part to an anchor-based approach. Additionally, improvements in semi-supervised learning mechanisms could enhance overall model efficiency. Regarding aligning high-level feature maps, we believe there is still considerable research potential.

Furthermore, our study has some limitations. Since the dataset used in this study is solely from one hospital, variations in equipment among different hospitals can lead to different feature distributions, potentially rendering the model less applicable to other hospital datasets. Future efforts could involve obtaining more diverse training data to enable the model to learn a broader range of lesion features, thereby enhancing its performance, robustness, and generalizability.